\documentclass[preprint,review,12pt]{elsarticle}

\usepackage{amssymb}
\usepackage[fleqn]{amsmath}
\usepackage{siunitx}

\usepackage{geometry}
\usepackage{graphicx}
\graphicspath{{figures/}}

\journal{Surface \& Coatings Technology}

\usepackage{tocloft}
\cftpagenumbersoff{figure}
\cftpagenumbersoff{table}

\begin{document}

\begin{frontmatter}

\title{Strain hardening-softening oscillations induced by nanoindentation in bulk solids}

\author[label1]{U.~Kanders \corref{cor1}}
\author[label2]{K.~Kanders}
\cortext[cor1]{Corresponding author, uldis.kanders@gmail.com}

\address[label1]{Institute of Solid State Physics, University of Latvia, Kengaraga Str. 8, Riga LV-1063, Latvia}
\address[label2]{Institute of Neuroinformatics, University \& ETH of Zurich, Winterthurerstr. 190, CH-8057 Zurich, Switzerland}

\begin{abstract}
Nanoindentation is a widely used method for sensitive exploration of the mechanical properties of micromechanical systems. We derived an empirical analysis technique to extract stress-strain field gradient and divergence representations from nanoindentation measurements. With this approach local gradients and heterogeneities can be discovered to obtain more detail about the sample's microstructure, thus enhancing the analytic capacity of the nanoindentation technique. We analysed nanoindentation tests of bulk solid substrates, namely bearing and tooling steel, and fused silica. Oscillations of the stress-strain field gradient and divergence induced in the subsurface layer by the nanoindentation experiment were revealed. The oscillations were especially prominent in single measurements at low indentation depths ($<$ 100 nm), whereas they were concealed in the averaged datasets. The amplitude of stress-strain field divergence oscillations decayed as a sublinear power-law when the indenter approached deeper atomic layers, with an exponent -0.9 for the steel and -0.8 for the fused silica. The oscillations were interpreted as alternating strain hardening-softening cycles induced in the subsurface layers under the indenter load. A selective assessment of elastic and plastic stress-strain field components indicated an elastic-plastic deformation process where the normal strain is transformed into the shear strain leading to a plastic deformation.
\end{abstract}


\end{frontmatter}

\section{Introduction}

\begin{figure}[h]
\centering
\includegraphics[width=0.6\linewidth]{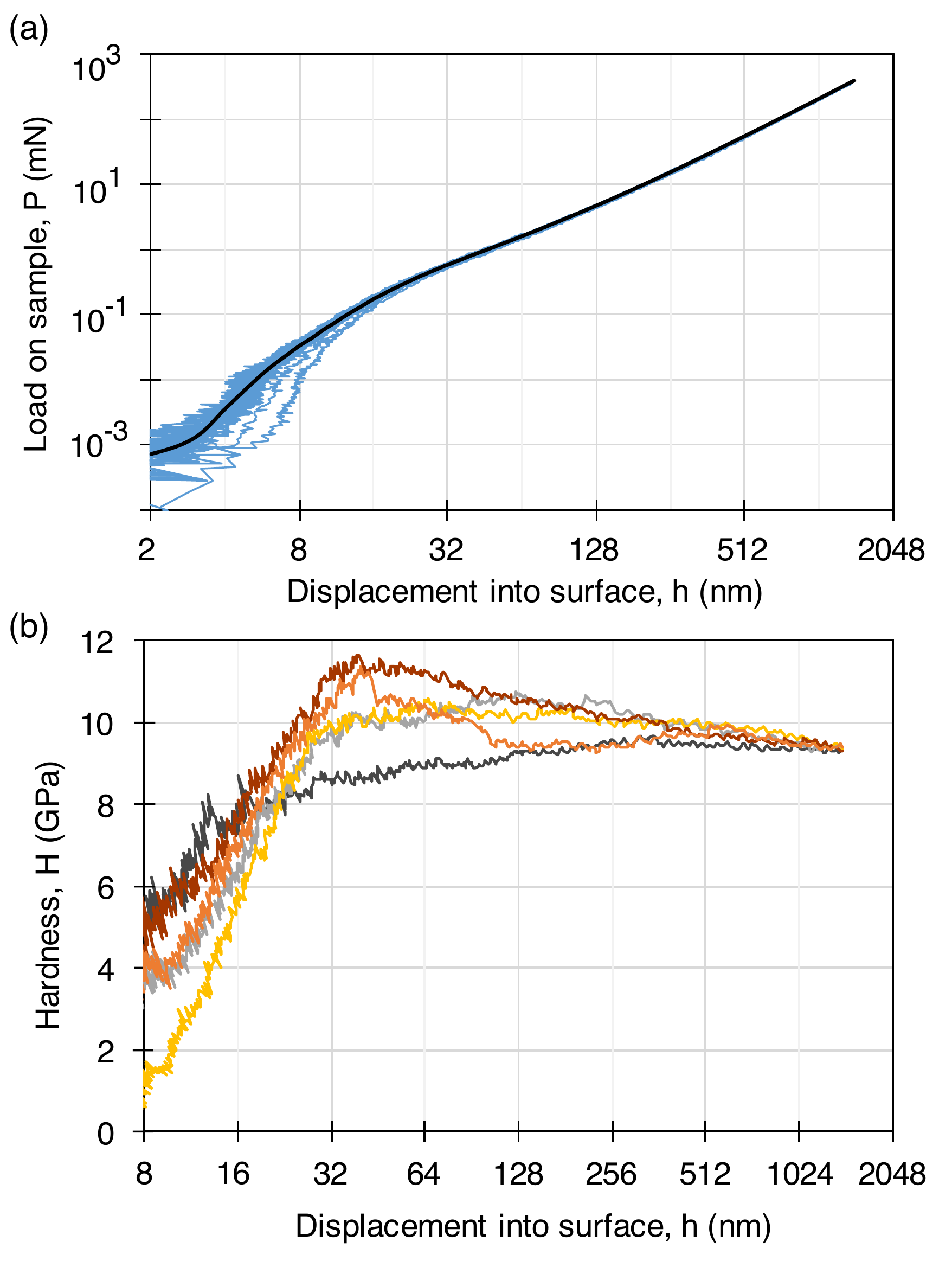}
\caption{
Variability in nanoindentation measurements: \textbf{(a)} Overlaid load-displacement curves from $n=19$ single measurements of the CZS steel sample studied in this article (blue, see text for the experimental details) and the averaged load-displacement curve (black); \textbf{(b)} Examples of single hardness measurements demonstrating that subsurface layer plasticity is locally rather different, but it converges to bulk hardness as the displacement into surface, $h$, increases. Note the use of logarithmic axis for $h$ to highlight the nanoindentation response at shallow penetration depths.
}
\label{Fig1}
\end{figure}

Nanoindentation is a powerful experimental technique to characterise the mechanical properties of small volume samples, such as thin films, subsurface layers of bulk solids or biological materials like bone, tooth enamel and even viruses \cite{Fisher2004, Oyen2009, Guo2005, Warren2006, Michel2006}. The measurement usually has high variability at shallow penetration depths (see Fig.~\ref{Fig1}). Accordingly, the material's hardness and elastic modulus is commonly calculated from data averaged over around ten or more single indentation tests at depths exceeding at least 200--300 nm. Preferably the measured variable has then approached some stable steady state value that corresponds to the so-called material property ``in bulk", often measured by the micro- or macroindentation techniques. However, there are also situations when a steady state nanoindentation response at increasing penetration depth cannot be generally achieved or expected. For example, it is not always possible to observe a stationary plateau in the nanoindentation measurement of hardness or elastic modulus in thin film samples when a strong reverse indentation size effect is present \cite{Sangwal2000, Kanders2015} or when the sample's substrate is influencing the experiment \cite{Saha2002, Manika2008}. In the context of using bulk solids as substrates for deposition of thin and very thin films, a detailed understanding of the substrates' surface and subsurface layers at depths of several hundred to a thousand nanometres is a prerequisite for an adequate assessment of the deposited thin films' mechanical properties. The subsurface layer of bulk solids, especially of polished steel substrates, is usually highly heterogeneous and would not yield a stable value of the apparent hardness or elastic modulus \cite{Guo2005}. In all these cases it would be valuable to probe the local gradients and inhomogeneities in a better detail to reveal more information about the microstructure of the sample.

In this article we derived a simple approach to extract the local stress-strain field (SSF) gradient and divergence representations from the nanoindentation experiment dataset. The strain-gradient representations, in principle, allow to discover weak structural heterogeneities, indicating, for example, interfaces between mechanically distinct local microzones within the sample or work hardening and softening processes induced underneath the indenter. We applied the derived strain gradient-divergence approach in the analysis of nanoindentation response of bulk solids commonly used as substrates for thin film deposition: different types of bearing and tooling steel as well as silica as a reference sample without a pretreated surface. A highly dynamic process taking place in the subsurface layer of the bulk solids during a nanoindentation experiment was revealed: oscillations of the stress-strain field gradient and divergence. The gradient-divergence oscillations were best discernible in the single measurement data at penetration depths below 100 nm, where the measurement has the highest variability and thus is commonly discarded. We associate these oscillations with alternating cycles of strain hardening and strain softening processes induced during the indentation experiment. To the best of our knowledge, such a phenomenon has not been reported previously.

\section{Methods}

\subsection{Derivation of the stress-strain field gradient and divergence representations}

In the nanoindentation experiment, each new indentation increment causes a mechanical shock that generates an elastic-plastic deformation wave which propagates throughout the sample and fades when the system is relaxed. As a result, a stress-strain tensor field of restoring forces is induced periodically in the sample. In this section we propose that useful information about the SSF can be extracted from the nanoindentation measurement, namely the representations of SSF gradient and divergence. The analysis technique derived below builds upon the strain gradient plasticity theory where the stress-strain relationship at any given point is considered in the context of deformation events in the long range vicinity of that point \cite{Fleck1997, Gao1999}.

In the expanding cavity model a heavily deformed hydrostatic core encases the indenter tip as deep and wide as the contact radius, $a_c$. The hydrostatic core is surrounded by superimposed hemispherical plastic and elastic deformation zones \cite{Fisher2004, Johnson1970}. The contact radius differs from penetration depth, $h$, by a constant factor, e.g., about 3 times ($a_c \approx 3h$) in the case of the sharp Berkovich indenter. However, when using logarithmic coordinates instead of linear ones for the penetration depth, the choice between log($h$) or log($a_c$) only shifts the function plot along $h$-axis and does not change its form. This and the fact that the peak pressure between the core and the plastic deformation zone is quite diffuse allows us to simplify the analysis of the SSF and use $h$ as an independent variable instead of $a_c$. The hydrostatic core is considered as an incompressible, homogeneous extension of the indenter tip that transduces the applied test load, $P(h)$, into the SSF. The indented system resists to further elastic-plastic deformation and compensates the applied test load according to Newton's third law. There is no such probe to measure directly the SSF components at each point $R(z \geq h)$  throughout the sample during a nanoindentation experiment, but one can measure the integrated echo of the SSF on the interface between the core and the plastic deformation zone $R(z=h)$ as the restoring force $F(h)= -P(h)$, or the total stress $\sigma_t(h) = F(h)/(2 \pi h^2)$. Note that $\sigma_t(h)$ can be interpreted as the potential energy density function of SSF. This allows us to introduce a generalized quantity, the SSF potential function $U(h)$ that is proportionally related to the stored potential deformation energy. Application of the nabla operator to $U(h)$ creates a gradient field $\nabla U(h)$. Strong gradient of the SSF at a point $R(z=h)$ is an evidence that the indented system is highly heterogeneous in a short range vicinity of this point, whereas a weak gradient is a good sign that the system is homogeneous even in a long range vicinity. We define the normalized gradient of the generalized potential function $U(h)$ as
\begin{equation}
grad(U(h)) \equiv U'(h) \equiv \frac{h}{U(h)}\nabla U(h).
\label{eqn_gradient}
\end{equation}
The gradient $U'(h)$ is a very simplified force vector field which represents to some extent the much more complex actual stress-strain tensor field beneath the loaded indenter. In turn, we define the SSF divergence as the divergence of the $U'(h)$ vector field by
\begin{equation}
div(U'(h)) \equiv U''(h) \equiv \nabla U'(h).
\label{eqn_divergence}
\end{equation}
Divergence is closely related to stress-strain field flux density -- the amount of stress-strain entering or leaving a given point. $U''(h)$ tells us at which h values the interface between the hydrostatic core and the plastic deformation zone acts as a stress-strain flux source or sink. It is easy to see that positive divergence means that the interface acts as a stress-strain flux source resulting in strain hardening effect, whereas negative divergence means that the interface acts as a stress-strain flux sink resulting in strain softening effect. 

The total stress, $\sigma_t$, can be broken down into elastic, $\sigma_e$, and plastic, $\sigma_p$, components, and each of these can be further separated into normal and shear stresses. Therefore, we can describe the total stress or the total potential energy density of the stress-strain field as a superposition of the elastic normal, $\sigma_{en}$, the elastic shear, $\sigma_{e\tau}$, the plastic normal, $\sigma_{pn}$, the plastic shear, $\sigma_{p\tau}$, stress components. It remains to be shown how to link these stress components to the appropriate experimental datasets obtained during a nanoindentation experiment. Knowing the surface area of the interface between the hydrostatic core and the plastic deformation zone, one can calculate the average total stress:
\begin{equation}
\sigma_t(h) = \frac{F(h)}{2\pi h^2} = \frac{-P(h)}{2\pi h^2} \Rightarrow \sigma_t(h) \propto \frac{-P(h)}{h^2}.
\label{eqn_total}
\end{equation}
We assume that the elastic and plastic stress fields are superimposed and have a joint interface with the core \cite{Johnson1985}. The elastic total stress, $\sigma_e(h)$, can be evaluated using the harmonic contact stiffness, $S(h)$, experimental dataset:
\begin{equation}
\sigma_e(h) = \frac{S(h)h}{2\pi h^2} = \frac{S(h)}{2\pi h} \Rightarrow \sigma_e(h) \propto \frac{S(h)}{h}.
\label{eqn_elastic}
\end{equation}
Elastic normal stress component is represented by the well-known relationship containing elastic modulus, $E(h)$: $\sigma_{en}(h)=E(h) \varepsilon_{en}$, where $\varepsilon_{en}$ is the elastic normal strain. In the nanoindentation experiment the increment $\delta h$ is changed progressively so that the incremental strain, $\varepsilon = \delta h/h$, is usually kept almost constant. Step by step indentation with a constant incremental strain has the advantage of logarithmically scaling the data density so that there are equal amounts of data at low and high loads. Therefore, we can treat the incremental strain as a constant variable and simplify to
\begin{equation}
\sigma_{en}(h) \propto E(h).
\label{eqn_elasticnormal}
\end{equation}
The plastic total stress component can represented by hardness, $H(h)$:
\begin{equation}
\sigma_{p}(h) \propto H(h).
\label{eqn_plasticnormal}
\end{equation}

Using the definition of SSF gradient from Eq. \ref{eqn_gradient}, we derive the specialized elastic-plastic strain gradients from Eqs. \ref{eqn_total}--\ref{eqn_plasticnormal}:
\begin{equation}
\begin{aligned}
&P'(h) \equiv \bigg ( \frac{h}{P(h)} \bigg ) \bigg( \frac{dP(h)}{dh} \bigg ) - 2, \\
&S'(h) \equiv \bigg ( \frac{h}{S(h)} \bigg ) \bigg( \frac{dS(h)}{dh} \bigg ) - 1, \\
&E'(h) \equiv \bigg ( \frac{h}{E(h)} \bigg ) \bigg ( \frac{dE(h)}{dh} \bigg),\\
&H'(h) \equiv \bigg ( \frac{h}{H(h)} \bigg ) \bigg ( \frac{dH(h)}{dh} \bigg ),
\end{aligned}
\end{equation}
where $P'(h)$ represents total strain gradient, $S'(h)$ represents elastic total strain gradient, $E'(h)$ represents elastic normal strain gradient, and $H'(h)$ represents plastic total strain gradient induced beneath the indenter. In the rest of the text we will refer to these functions as the corresponding strain gradients instead of strain gradient representations because they differ by a constant factor only. In an analogous way we also derive the specialized total, $P''(h)$, elastic total, $S''(h)$, elastic normal, $E''(h)$, and plastic total, $H''(h)$ divergences using the definition in Eq. \ref{eqn_divergence}:
\begin{equation}
\begin{aligned}
&P''(h) \equiv \nabla P'(h),\\
&S''(h) \equiv \nabla S'(h),\\
&E''(h) \equiv \nabla E'(h),\\
&H''(h) \equiv \nabla H'(h).\\
\end{aligned}
\end{equation}

In practice we estimated the gradient and divergence representations by calculating the analytic derivative of a polynomial fit to the measurement data within a sliding window of $2m+1$ points. Results with 1-st (i.e., linear) and 2-nd order fits were already found to be satisfactory, with $m =$ 8, 9 or 10 providing the optimal window lengths indicated by robust accuracy of the fits (generally, R-squared value $\approx$ 0.90). The differences between the linear and 2-nd order fit were less than 2\% of peak amplitude at very shallow penetration depths ($h<32$ nm). In this report we will present results obtained using a linear fit with $m=8$. The exact length of the window did not affect the results strongly: a slightly smaller $m$ value yielded noisier gradient and divergence representations, whereas a slightly larger $m$ rendered them smoother. In general, the sliding window should not be too large as, otherwise, the fine structural features of the gradient and divergence representations will be smoothened out.

\subsection{Experimental details}

Various bulk samples of steel, glass and fused silica substrates were tested in the nanoindentation experiments. The steel samples, labelled as 100Cr6 (OTAI Special Steel) and SCH (Schaeffler Technologies AG) were classified as bearing steel. The other category of steel samples labelled as CZS and HVG (ProfProkat) corresponded more to tooling steel. All of them were of high quality chrome low alloy steel closely matching the AISI 52100 steel with minor modifications depending on their respective manufacturers. The glass sample was the hard glass microscope slide VB5 656; the fused silica sample was the Corning HPFS 7980, Standard Grade high purity synthetic amorphous silicon dioxide.

Surface morphology, microstructure, phase structure and elemental composition of the samples was estimated by scanning electron microscope (model Hitachi S-4800) equipped with the energy dispersive spectrometer B-Quantax and X-ray fluorescence spectrometer S4 PIONEER. X-ray diffraction structure measurements were performed by Bruker D8 Advance.

\begin{figure}[h]
\centering
\includegraphics[width=0.6\linewidth]{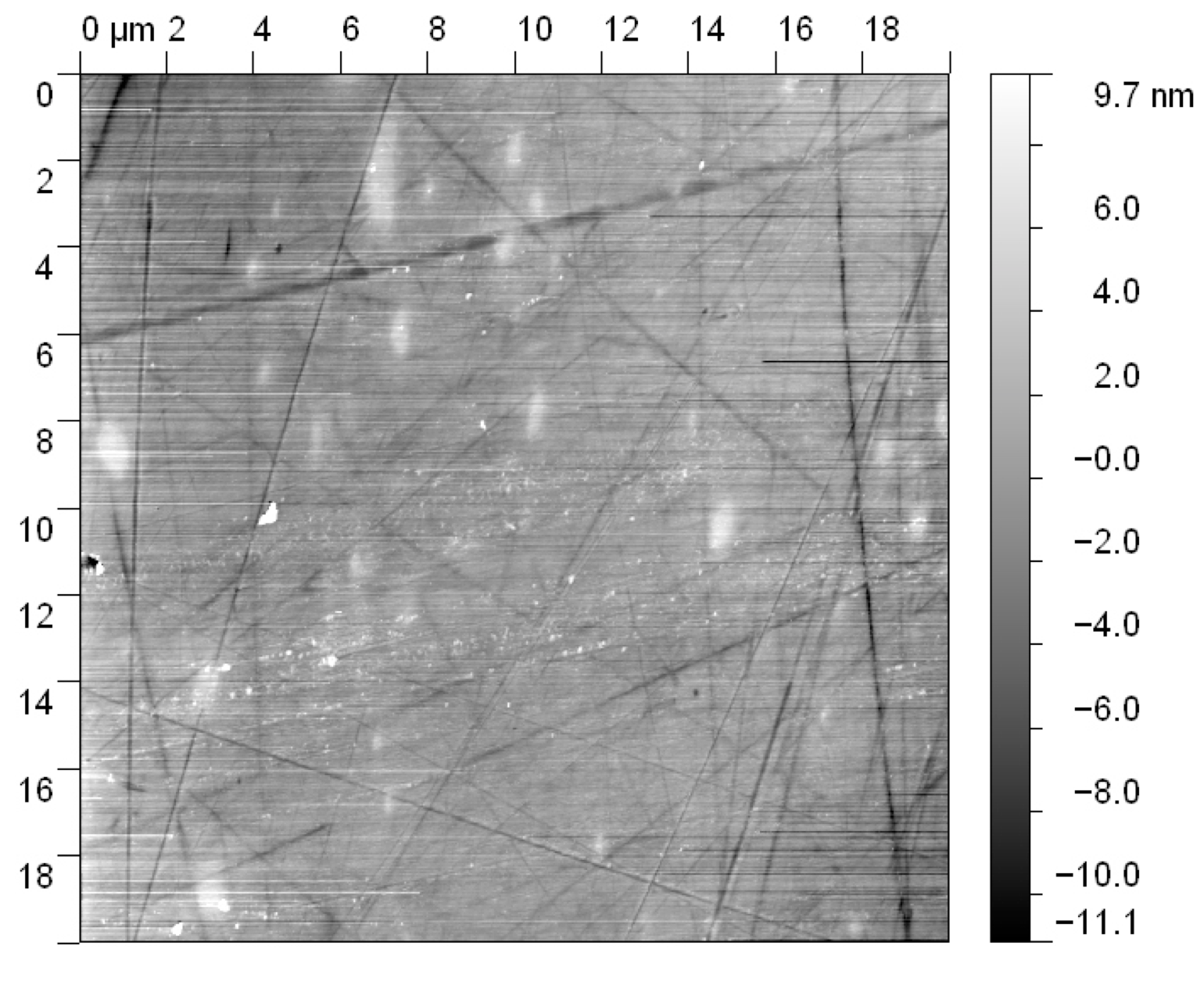}
\caption{
Characterisation of the CZS steel sample using AFM: surface RMS roughness measurement of $R_a=1.5$ nm, $RMS=2.2$ nm.
}
\label{Fig2}
\end{figure}

Surface root mean square (RMS) roughness was measured by atomic force microscope (AFM, model Veeco CP-II and Asylum Research, model MFP-3D) in tapping mode. The surface micrographs and phase images were taken from surface area of 20 \SI{}{\micro\metre} $\times$ 20 \SI{}{\micro\metre} (see Fig.~\ref{Fig2} for an example of the CZS sample's surface). Surface profiler Dektak 150 was used to probe surface texture and roughness over a wider area, on the order of mm. The Dektak 150 resolution features enabled precise planarity scans for measuring radius of curvature, flatness, and waviness.

Instrumented depth sensing nanoindentation experiments were performed by G200 Nano Indenter (Agilent, USA) with a sharp Berkovich diamond indenter (tip radius $<$ 20 nm). Measurements were made in the continuous stiffness measurement (CSM) mode \cite{Hay2010} and in the BASIC mode at different values of the maximum load. The load capability of the Nano Indenter G200 can reach up to 600 mN with the standard option. The measurements in the CSM mode were carried out in a load-controlled manner where the logarithmic indentation strain rate $(dh/dt)/h$ was kept constant at the level of 0.05 s$^{-1}$ and the superimposed harmonically oscillating force frequency was 45 Hz. The nanoindenter was calibrated using a reference sample of fused silica. The hardness and elastic modulus of the samples were calculated by the MTS TestWorks 4 software using Oliver-Pharr method \cite{Oliver2004}. 

\section{Results and discussion}

We performed nanoindentation experiments on steel substrates labelled in our experiments as CZS, 100Cr6, SCH and HVG. With regard to the chemical composition and heat treatment all of these samples were close to the AISI 52100 steel. As silicon wafers and glass slides are often used as substrates for thin film systems, they were studied in nanoindentation experiments as well. SSF gradient and divergence oscillations induced by nanoindentation were observed in all of the materials mentioned above and they manifested similar strain hardening-softening cycles when the sharp Berkovich indenter penetrated the subsurface layer. For the clarity of the discussion we focus on reporting a few typical results of the nanoindentation measurements of the CZS sample, which are representative of all the steel samples examined in our investigation. In addition, results from the well known and widely used reference material fused silica are also displayed to showcase the generality of the SSF gradient-divergence oscillation phenomenon. We also provide a preliminary quantitative analysis of the oscillations, but we leave a more detailed analytic treatment outside the scope of this report.

\subsection{SSF gradient and divergence oscillations}

\begin{figure}[h]
\centering
\includegraphics[width=\textwidth]{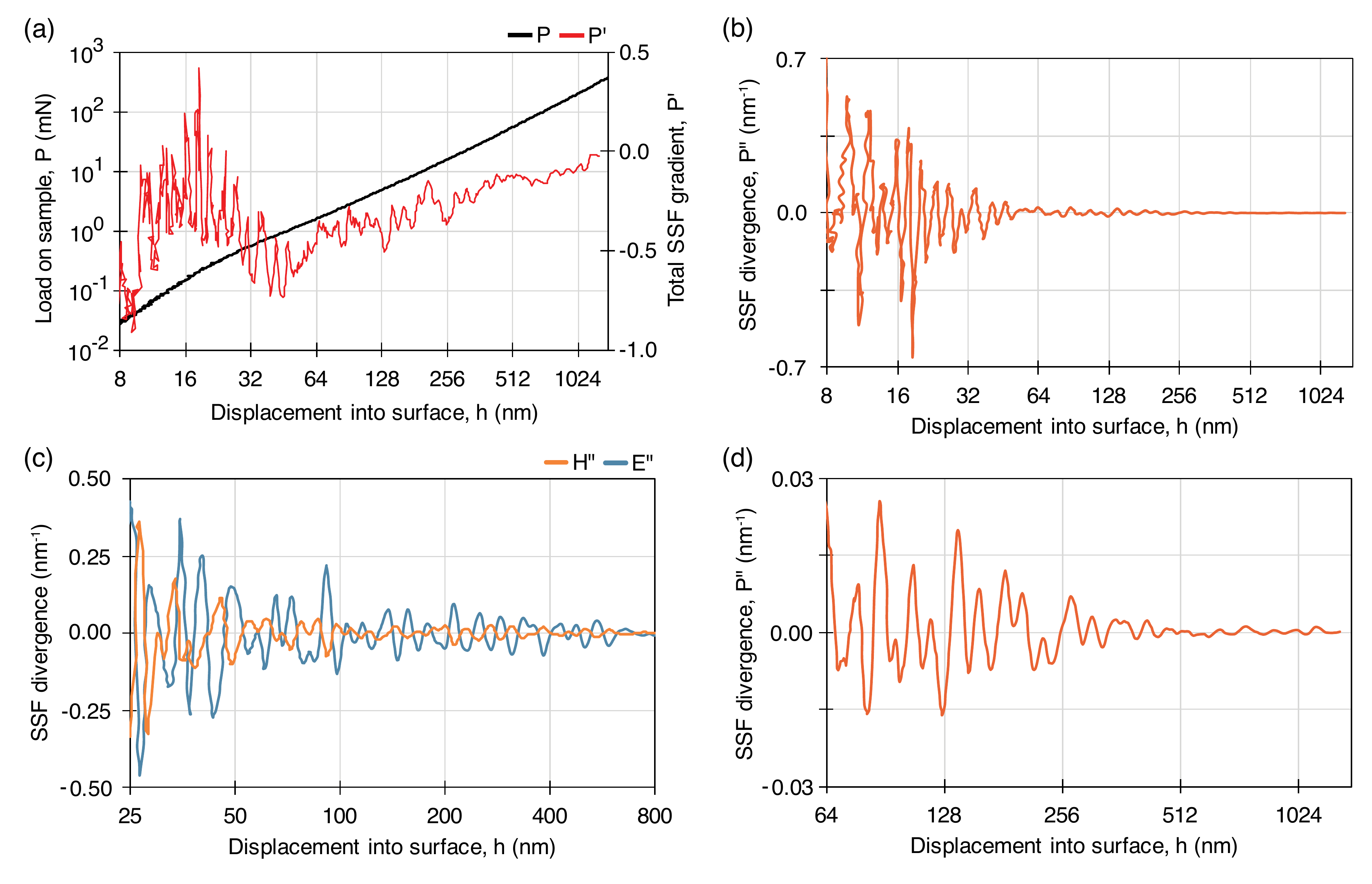}
\caption{Total stress-strain field gradient and divergence oscillations in the CZS steel sample: \textbf{(a)} A typical example of the load-displacement, $P$-$h$, curve (black) and the total SSF gradient, $P'$-$h$, curve (red) obtained from a single measurement in the CSM mode; \textbf{(b)} Total SSF divergence oscillations obtained from the same measurement; \textbf{(c)} Normal plastic ($H''$-$h$ curve, orange) and normal elastic ($E''$-$h$ curve, blue) SSF divergence oscillations calculated from the same measurement; \textbf{(d)} Close-up of the total SSF divergence low amplitude oscillations from (b) at greater penetration depths.}
\label{Fig3}
\end{figure}

Fig.~\ref{Fig3}a shows a typical example of a single nanoindentation test measurement (as opposed to an averaged measurement) in the CSM mode of the CZS steel sample. The load-displacement curve is contrasted with the corresponding total SSF gradient curve which was derived using the approach outlined in Methods. Note that we use logarithmic axis for the displacement into surface, $h$, to highlight nanoindentation responses at the subsurface layers. Clear oscillations of the SSF gradient can be observed already starting from 8 nm depth despite the measurement noise. The total SSF divergence oscillations are shown in Fig.~\ref{Fig3}b. Very similar oscillations were observed for all 19 single nanoindentation tests of the CZS steel sample. The SSF gradient-divergence oscillations were observed throughout the indentation measurement, even at large indenter penetration depths (see Fig.~\ref{Fig3}d, where the SSF divergence oscillations are present at up to more than 1000 nm). Therefore, they cannot be solely attributed to the structural inhomogeneities related to surface roughness, which based on the RMS measurements was only 2--3 nm (see Fig.~\ref{Fig2}), or the superficial layer.

The oscillations of the strain gradient might be caused by alternating normal-shear strain transformation processes converting strains into each other: as the sharp Berkovich indenter penetrates the sample the normal strain builds up until the sample's structure cannot withstand it anymore and some sliding of elastically deformed region occurs, i.e. shear strain leads to plastic deformation. Furthermore, the incremental stresses induced by indenter interact with the structural stresses inherent in the subsurface layer. The amplitude of the SSF gradient-divergence oscillations was larger closer to the surface, where subsurface layer acts as a large structural defect in comparison to the atomic structure in bulk. Subsequently the amplitude decayed rather rapidly as the indenter penetrated deeper into the sample and approached its bulk structure that is far enough from the surface.

Fig.~\ref{Fig3}c shows the plastic total SSF divergence, $H''-h$ curve, and the elastic normal SSF divergence, $E''-h$ curve, that were derived from the hardness and elastic modulus indentation datasets, respectively. The intervals where the divergence of the corresponding component of the SSF is positive can be interpreted as strain hardening of the sample's subsurface layer. Within these depth intervals the subsurface layer acts as a SSF flux source. Similarly, the penetration depth intervals of negative divergence can be interpreted as strain softening of the subsurface layer. The oscillations in Fig.~\ref{Fig3}c reveal a structure of strain hardening-softening zones induced by the sharp Berkovich indenter. The oscillations of the elastic and plastic SSF divergences were out of phase, which indicates that elastic and plastic deformation processes in the subsurface layer happened in an alternating fashion. This is in a good agreement with the physical nature of the elastic-plastic deformation. The sample in the subsurface layer is expected to be rather heterogeneous and, therefore, nanovoids can be present, which then could be packed by the applied load during the nanoindentation measurement. The process of packing is largely an elastic deformation, which is indicated by the positive divergence of the elastic normal SSF divergence ($E''(h) > 0$). However, this can happen only up to the point when the elastic limit is exceeded. After the stress reaches the micro-yield point, a plastic deformation, i.e., a structural changes take place, which is indicated by the positive plastic total SSF divergence ($H''(h) > 0$).

\subsection{SSF gradient and divergence from averaged datasets}

\begin{figure}[h]
\centering
\includegraphics[width=0.6\linewidth]{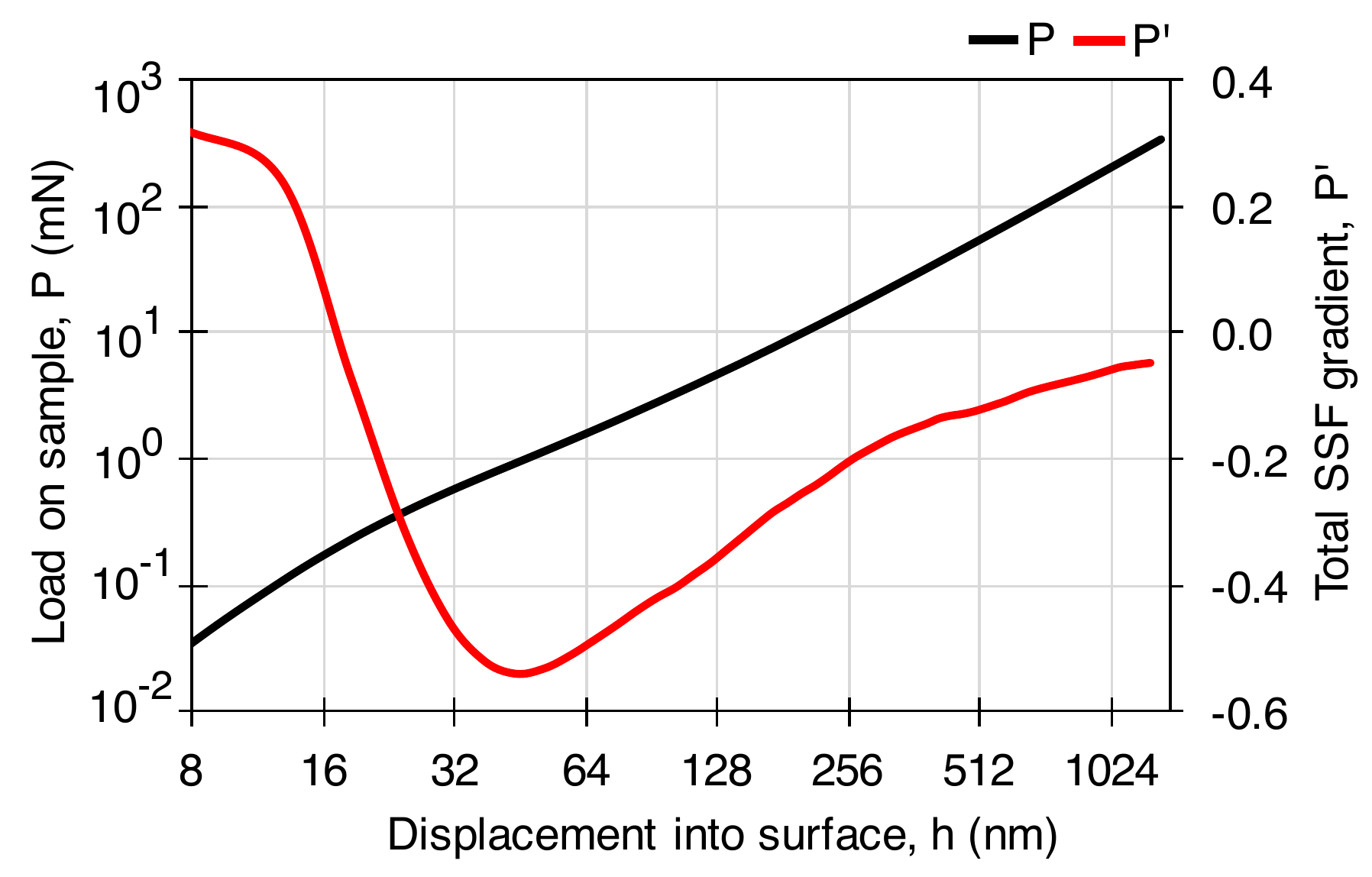}
\caption{
The strain gradient oscillations of the CZS steel sample are smoothened out when the total strain gradient-displacement, $P'$-$h$, curve (red) is calculated from the averaged ($n=19$) load-displacement, $P$-$h$ (black).
}
\label{Fig4}
\end{figure}

\begin{figure}[h]
\centering
\includegraphics[width=\linewidth]{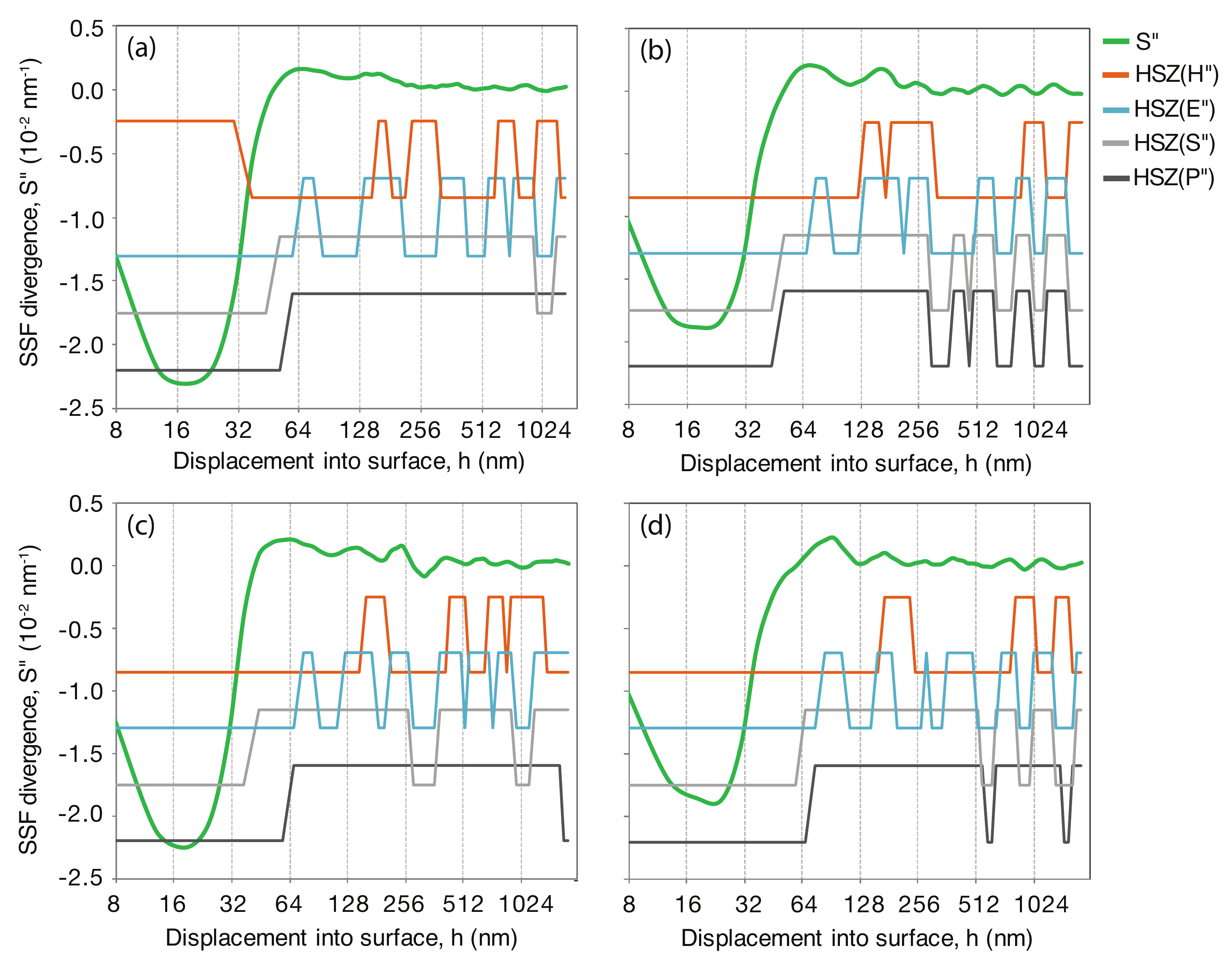}
\caption{
SSF divergence oscillations are also present, albeit much less prominently, in the averaged nanoindentation datasets of the steel samples: \textbf{(a)} CZS, \textbf{(b)} 100Cr6, \textbf{(c)} SCH, \textbf{(d)} HGV. Total elastic SSF divergence, $S''$-$h$, curve, is shown together with the strain harden-softening zones (HSZ) of the corresponding SSF divergence components; the HSZ curves are vertically offset for presentation purposes.
}
\label{Fig5}
\end{figure}

The SSF gradient calculated from the averaged load-displacement curves of the CZS sample was much smoother than in the single measurements and the oscillations were practically not visible (Fig. \ref{Fig4}). SSF divergence oscillations could still be detected albeit they were markedly smaller in amplitude compared to the single tests (Fig. \ref{Fig5}). The strain hardening-softening zones (HSZ) in Fig. 5 are highlighted by applying the sign function on the corresponding components of SSF divergence (e.g., $HSZ(E'') = sign(E'')$), revealing similar pattern seen in the case of single measurements: the strain hardening (HSZ $> 0$) and softening (HSZ $< 0$) cycles of the elastic and plastic components of SSF divergence tend to happen in an alternating fashion. However, generally the attempts to detect strain gradient-divergence oscillations from the averaged dataset will be impaired, because the oscillations of the single indentation tests are shifted in phase with respect to each other and averaging will to a large extent smoothen them out. However, less subtle fluctuations, e.g., due to a layered atomic microstructure in the case of thin films may still be detectable with the strain gradient-divergence approach also in the averaged indentation dataset \cite{Kanders2015}. For example, the averaged SSF gradient curve for the CZS steel had a pronounced convex shape (see Fig. \ref{Fig4}), which is in a good agreement with the model where the subsurface layer plays role of the skin of bulk materials and it has to endure a very large pressure difference between the outside and inside of the bulk sample (i.e., the skin effect). This is easy to understand as the pressure at the free surface of the sample is about 0.1 MPa, whereas the pressure inside the bulk samples reaches values of about 1000 MPa and higher. The pressure between the outside and inside of the sample differs on the order of 10$^4$--10$^5$ times, which makes the atomic layers within the subsurface area very stressed in comparison to those lying deeper inside the sample, far away from the free surface. The stresses within the subsurface layer are lowering when indenter is loading the surface of the bulk sample, which can be interpreted as strain softening of the subsurface layer at the beginning phase of the nanoindentation measurement.

\subsection{Additional tests for the presence of strain gradient oscillations}

\begin{figure}[h]
\centering
\includegraphics[width=0.6\linewidth]{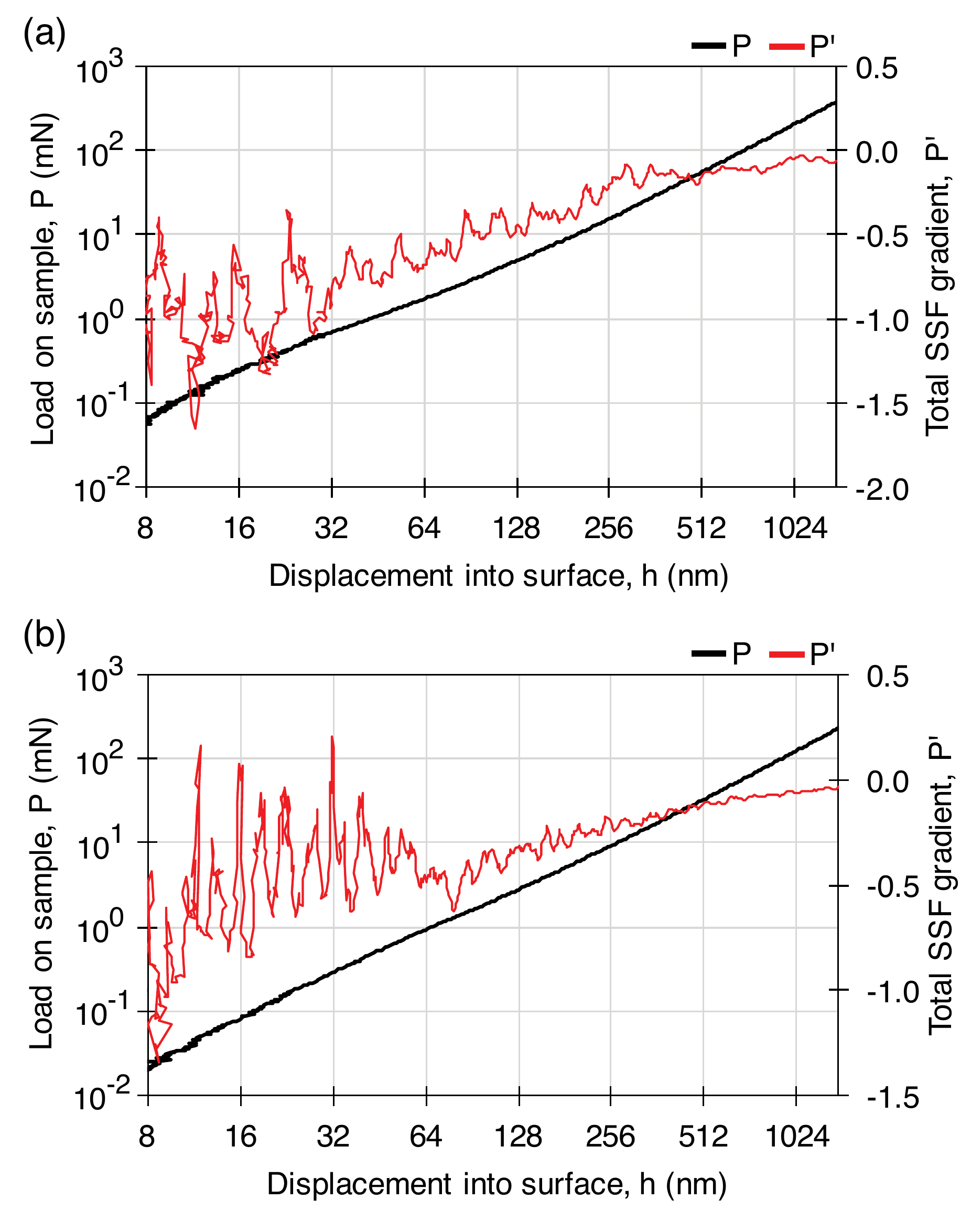}
\caption{
Additional tests of the strain gradient oscillation phenomenon: the load-displacement, $P$-$h$, curve (black) and the total SSF gradient, $P"$-$h$, curve (red) from a typical single measurement in \textbf{(a)} CZS steel in BASIC mode and \textbf{(b)} fused silica in CSM mode.
}
\label{Fig6}
\end{figure}

In the CSM mode weak superimposed oscillations are applied additionally to the load-time frame in order to measure the apparent hardness and elastic modulus throughout the loading process. This might in principle introduce some fluctuations of the sample's mechanical properties and potentially even alter the indentation measurement results \cite{Siu2013}. The CSM mode oscillations happen on a much faster timescale than the strain gradient-divergence oscillations and thus it was highly unlikely that CMS mode could affect the results. However, to confirm that the stress-strain gradient oscillations are not an artefact of the CSM loading mode, additional nanoindentation tests were made in the so-called BASIC mode, where the applied load is increasing up to the maximum load without any shallow oscillations during the loading process. Fig. \ref{Fig6}a shows that the stress-strain field gradient exhibits clear oscillations also when the indentation experiment is carried out in the BASIC mode.

Subsurface layer mechanics of steel samples may differ from that of glass samples because the surface of steel samples needs grinding and polishing before it is possible to carry out nanoindentation experiments. In turn, grinding and polishing of the steel sample may result in, e.g., a Beilby layer having somewhat specific structure and mechanics \cite{Beilby1921,Bhushan2001}. In contrast, glass samples in most cases do not need a special treatment of the surface after they are manufactured from the glass liquid phase. To test whether the SSF gradient oscillations are not just a peculiarity of the steel material, we indented a sample of a well-known reference material, the fused silica. Prominent stress-strain field gradient oscillations were present in all 8 single indentation tests we performed (see Fig. \ref{Fig6}b for a typical example).

\subsection{Quantitative features of the oscillations}

\begin{figure}[h!]
\centering
\includegraphics[width=0.6\linewidth]{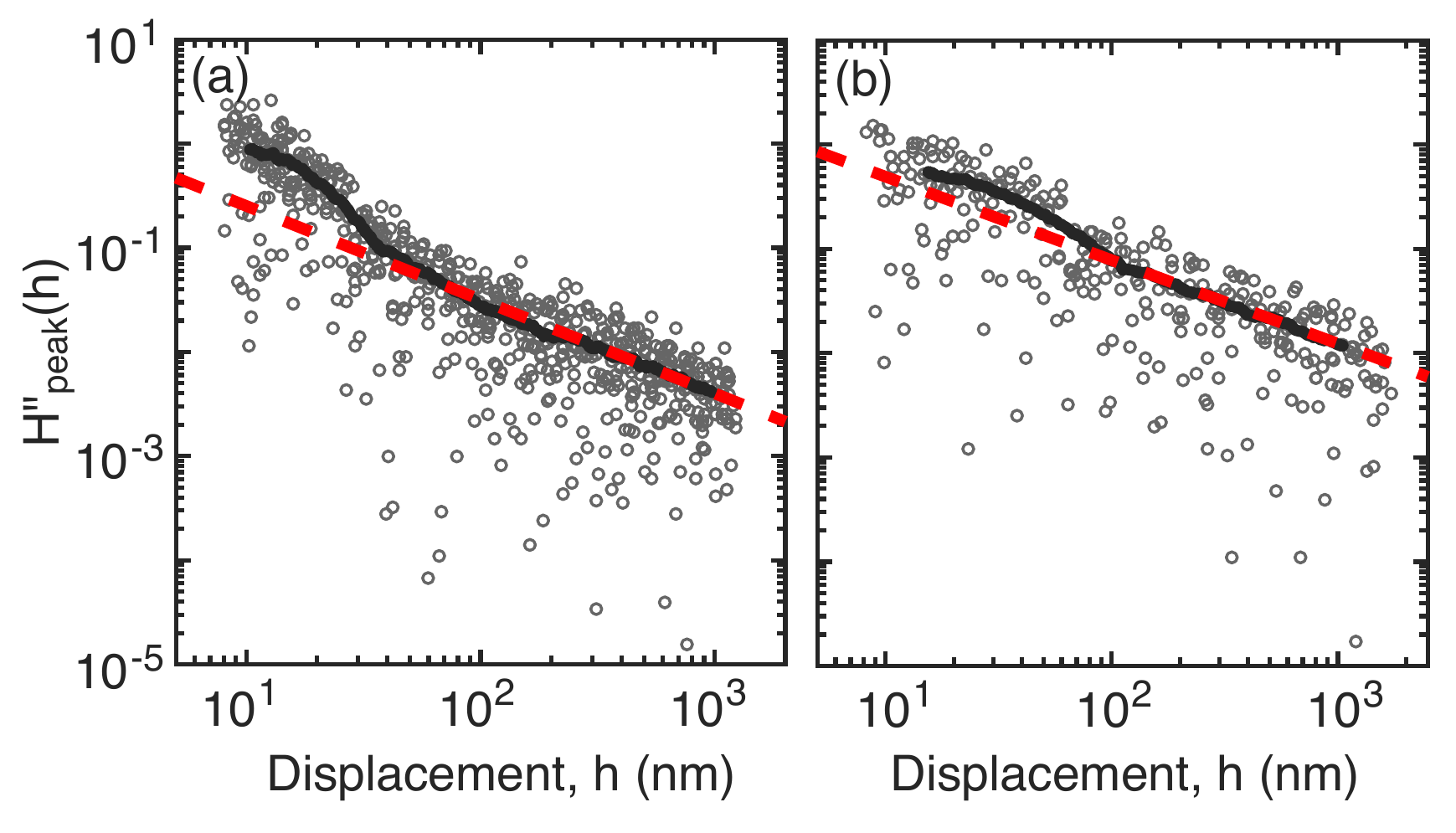}
\caption{
$H''(h)$ peak amplitude, $H''_{peak}(h)$, as function of the penetration depth, $h$, for \textbf{(a)} steel ($n=19$ tests) and \textbf{(b)} fused silica ($n=8$ tests). Gray open circles are pooled data from all tests; black, solid lines show a moving average over 60 points of the pooled and sorted dataset to reveal the trend of the $H''_{peak}(h)$ decay; red, dashed line demonstrates a power-law relationship $H''_{peak}(h) \propto h^\alpha$  with the exponent $\alpha=-0.9$ in (a) and $\alpha=-0.8$ in (b).
}
\label{Fig7}
\end{figure}

\begin{figure}[h!]
\centering
\includegraphics[width=0.6\linewidth]{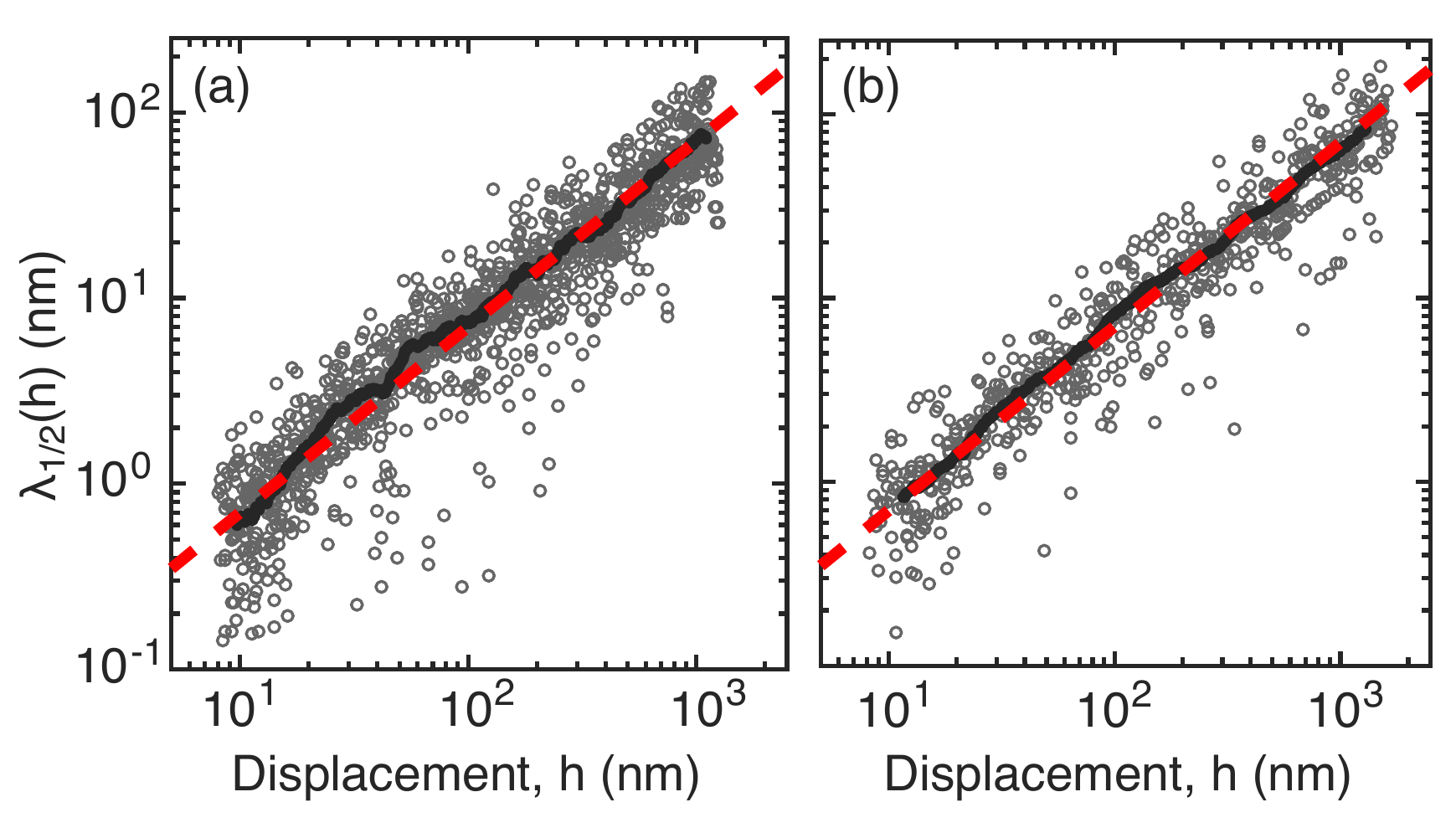}
\caption{
Half-wave length of the $H''(h)$  oscillations, $\lambda_{1/2}(h)$, as a function of the penetration depth, $h$, for \textbf{(a)} steel ($n=19$ tests) and \textbf{(b)} fused silica ($n=8$ tests). Gray open circles are pooled data from all tests; black, solid line shows a moving average over 60 points of the pooled and sorted dataset to reveal the trend of the $\lambda_{1/2}(h)$ growth; red, dashed line demonstrates a linear relationship $\lambda_{1/2}(h) \propto h$ in both (a) and (b).
}
\label{Fig8}
\end{figure}

Are the stress-strain field gradient-divergence oscillations specific for each material? We compared the decay behavior of the total plastic SSF divergence oscillations of CZS steel and fused silica by finding the positive peaks of $H''$-$h$ curves, $H''_{peak}(h)$,  and plotting them against the penetration depth, $h$ (Fig. \ref{Fig7}). $H''_{peak}(h)$ was defined as the maximum positive value between two consecutive crossings of the displacement $h$-axis of the $H''$-$h$ curve. The amplitude of oscillations for the steel sample was generally lower except at very shallow penetration depths. The amplitude for both materials decayed monotonically with increasing penetration depth, but there appear to be two different stages of oscillations: an initial stage of gradually decaying, higher amplitude oscillations that is followed by a rather sharp, discontinuous drop in amplitude at a penetration depth of approximately $h=30$ nm for the CZS sample and $h=60$ nm for the fused silica. After the sharp decrease, the amplitude again continues to decay more gradually. The smaller amplitude oscillations appear to decay to a first approximation as a power-law, i.e., $H''_{peak}(h) \propto h^\alpha$, with the exponent $\alpha = -0.9$ and $\alpha = -0.8$ for steel and fused silica, respectively. Interestingly, the half-wave length of the $H''(h)$ oscillations, $\lambda_{1/2}(h)$, defined as the distance in $h$ between two consecutive crossings of the $h$-axis, increased linearly with increasing penetration depth for both steel and fused silica (Fig. \ref{Fig8}). Alternatively, the half-wave length can be expressed as a function of the wave number, $N$, in which case $\lambda_{1/2}(N)$ grows exponentially. Overall, the analysis suggest that the SSF divergence oscillations have a quasi-regular, material specific structure with the amplitude and period changing in a regular manner, which follows a well-defined mathematical relationship with respect to the indentation depth. However, further investigations together with a more detailed morphological assessment techniques as a reference (e.g., using cross-sectional transmission electron microscopy \cite{Lloyd2005}) are necessary to elucidate oscillation features pertaining to specific mechanical properties of the sample's subsurface layer.

\section{Conclusions and outlook}

In this study a simple approach to assess stress-strain field (SSF) gradient and divergence from nanoindentation measurement data was derived and used to investigate several steel and glass samples, which were chosen as representative materials of bulk solids commonly used as substrates for thin film deposition. We found that nanoindentation experiment with a sharp Berkovich indenter induces SSF gradient-divergence oscillations in the bulk solids. The oscillations were especially prominent at low indentation depths ($<$ 100 nm) indicating that they are primarily a property of the bulk solid's subsurface layer. The oscillation amplitude decayed rapidly as the indenter approached deeper atomic layers, but some fluctuations could be still detected even at displacements greater than 1000 nm. It was crucial for the detection of the strain gradient-divergence oscillations to use the single measurement data: the oscillations were shifted in phase from one measurement to another and averaging essentially smoothened them out. Nevertheless, some fluctuations of SSF divergence could still be revealed even from the averaged data, albeit with a significantly reduced amplitude. To the best of our knowledge, such SSF gradient and divergence oscillations detected from nanoindentation measurements have not been previously reported.

We interpret the oscillations as alternating strain hardening and softening cycles taking place in the subsurface layers under the indenter load. The fading of the oscillations at deeper atomic layers of the sample (i.e., closer to the ``in bulk" region) manifest that deformation processes underneath the indenter have reached a steady state where the nanohardness and elastic modulus do not change anymore under continued loading of the sample. Although the elastic and plastic stress-strain fields are superimposed, it was possible to separate their respective components and study elastic and plastic deformation processes selectively. We observed that elastic and plastic strain hardening and softening cycles happen in an alternating fashion which is in agreement with the nature of an elastic-plastic deformation.

The strain hardening-softening oscillations were observed in all the studied samples. Steel has a crystalline lattice structure, whereas glass is an amorphous material, and the specific deformation mechanism in both types of materials could be different. However, these results suggest that the alternating cycles of work hardening and softening is a general phenomenon occurring in bulk solids when their surface is loaded by incremental stress. Preliminary quantitative analysis suggest that the oscillations have features that are specific to the different materials, but further investigations, preferably together with detailed morphological characterisation of the sample's microstructure are necessary to fully uncover the potentially valuable information about the mechanical properties of the material's subsurface layer hidden in the strain gradient-divergence oscillations.

The stress-strain gradient approach is not limited to the investigation of bulk solids. In fact, the most promising application would be in the studies of layered thin film structures, where the assessment of the mechanical properties is especially challenging \cite{Kanders2015, Misra1998, Maniks2015}. The influences of the surface and substrate are present throughout the thin film sample, which can result in indentation size effects that preclude a precise estimation of the true hardness and true elastic modulus. Furthermore, the thin films can be highly heterogeneous with gradients of changing chemical composition throughout the sample. In such cases, the strain gradient representations can be used to define an analytic criterion for determining the indentation depth at which the apparent values are the closest to the true values: the true mechanical properties should be read out in the most homogeneous regions within the sample, i.e., at the penetration depth where the strain gradient is the closest to zero. Stress-strain gradient representations can also inform about the microstructure of the thin film by revealing a pattern of strain gradient peaks and valleys which could indicate interfaces between sub-layers of the thin film and relative ``in-bulk" zones within the thin film structure, respectively,. Such analyses can be performed with the averaged datasets (cf. Ref. \cite{Kanders2015}), but we encourage the use of single measurements, which can provide much more detailed information.

\section{Acknowledgments}

The authors thank Dr. Roberts Zabels at the Institute of Solid State Physics, University of Latvia for providing the datasets of the nanoindentation and AFM experiments, and Prof. Janis Maniks for encouraging and helpful discussions.


\end{document}